\documentclass[format=acmsmall, review=false, screen=true]{acmart}
\pdfoutput=1
\usepackage{url}
\usepackage{booktabs} 
\usepackage{graphicx}  
\usepackage{amsmath}
\usepackage{makecell}

\setcopyright{acmlicensed}
\acmJournal{PACMHCI}
\acmYear{2018} \acmVolume{2} \acmNumber{CSCW} \acmArticle{43} \acmMonth{11} \acmPrice{15.00}\acmDOI{10.1145/3274312}

\received{April 2018}
\received[revised]{July 2018}
\received[accepted]{September 2018}

\begin{document}
\title[Coloring in the Links]{Coloring in the Links: Capturing Social Ties as They are Perceived}

\author{Sebastian Deri}
\affiliation{%
  \institution{Cornell University}
  \city{Ithaca}
	\state{NY}
	\country{USA}
}
\email{sebastian.deri@gmail.com}

\author{J{\'e}r{\'e}mie Rappaz}
\affiliation{%
  \institution{EPFL}
	\city{Lausanne}
  \country{Switzerland}
}
\email{jeremie.rappaz@epfl.ch}

\author{Luca Maria Aiello}
\orcid{0000-0002-0654-2527}
\affiliation{%
  \institution{Nokia Bell Labs}
  \city{Cambridge}
  \country{United Kingdom}
	}
\email{luca.aiello@nokia-bell-labs.com}

\author{Daniele Quercia}
\affiliation{%
  \institution{Nokia Bell Labs}
	\city{Cambridge}
  \country{United Kingdom}
	}
\email{quercia@cantab.net}

\begin{abstract}
The richness that characterizes relationships is often absent when they are modeled using computational methods in network science. Typically, relationships are represented simply as links, perhaps with weights. The lack of finer granularity is due in part to the fact that, aside from linkage and strength, no fundamental or immediately obvious dimensions exist along which to categorize relationships. Here we propose a set of dimensions that capture major components of many relationships -- derived both from relevant academic literature and people's everyday descriptions of their relationships. We first review prominent findings in sociology and social psychology, highlighting dimensions that have been widely used to categorize social relationships. Next, we examine the validity of these dimensions empirically in two crowd-sourced experiments. Ultimately, we arrive at a set of ten major dimensions that can be used to categorize relationships: similarity, trust, romance, social support, identity, respect, knowledge exchange, power, fun, and conflict. These ten dimensions, while not dispositive, offer higher resolution than existing models. Indeed, we show that one can more accurately predict missing links in a social graph by using these dimensions than by using a state-of-the-art link embeddedness method. We also describe tinghy.org, an online platform we built to collect data about how social media users perceive their online relationships, allowing us to examine these dimensions at scale. Overall, by proposing a new way of modeling social graphs, our work aims to contribute both to theory in network science and practice in the design of social-networking applications.
\end{abstract}

%
%



\maketitle

\renewcommand{\shortauthors}{S. Deri et al.}

\section{Introduction}\label{sec:intro}

Social relationships are paramount. They determine where we work, who we marry, and what we do. Their role is no less central online, where social-networking websites account for a large portion of the time people spend on the internet~\cite{pew2016}. In 2017, for example, Facebook was the third most visited website in the world, behind only Google and Youtube~\cite{alexa2017}; and, on average, users were estimated to spend almost an hour per day on the site~\cite{stewart2016}.

Unsurprisingly, researchers have spent a great deal of time trying to capture and model social relationships both online and offline~\cite{aiello17nature}. One of the dominant paradigms that has emerged is to think of and model social relationships in terms of tie strength~\cite{Granovetter1973strength}. Such approaches have no doubt been fruitful and certainly capture a key element of social relationships. For example, weak ties with acquaintances are critical in finding a job or generating creative ideas, while strong ties with family or close friends provide crucial emotional and social support. Tie strength is even predictive of subjective well-being~\cite{burke16}.

Yet, the focus on tie strength as a way of understanding and modeling social relationships has its limitations. Not all ties of the same strength are created equal~\cite{wellman90different}. Weak ties can be the way through which new job opportunities are discovered by professionals but not by blue collar workers; for people of low status, weak ties of a similar status are not generally useful or far reaching. Meanwhile, strong ties with parents offer financial support but rarely companionship, while strong ties with friends offer the opposite. Many other factors such as trust, power, and group affiliation play an important role in social relationships and are not captured by tie strength alone.

In this paper, we develop a simple way for incorporating more of this social richness and nuance into social network techniques. We first review relevant literature in sociology and social psychology from which we obtain 8 tentative dimensions along which relationships could be classified (\S\ref{sec:meth_lit}). Independently, we then ask 200 crowd-sourcing users to describe their relationships using everyday words and then to rate these words through a structured survey (\S\ref{sec:meth_st1}). Based on their responses, we add two new dimensions to the previous eight. We then apply our final ten dimensions to a link prediction task (\S\ref{sec:linkprediction}), in which an improvement in prediction accuracy suggests that these dimensions capture important aspects of a network's structure. In the final discussion (\S\ref{sec:discussion}), we compare our categorization to existing multidimensional relationship models and present \url{www.tinghy.org}, a web platform that allows users to label their online relationships with our dimensions, resulting in a large-scale collection of relationship labels.

\section{Related Work}\label{sec:related}

Attempts to model social relationships are of course not new. So, we begin by summarizing previous approaches to modeling social relationships in the study of tie strength, online conversations, multiplex networks, and multi-dimensional models in sociology---and we explain how our effort differs from each.

\subsection{Tie strength}

Since its inception, research in social network analysis has focused on modeling networks and social systems with simple nodes and edges~\cite{wasserman94social,newman03structure}. These abstract models were soon extended to incorporate edge weight~\cite{barrat04architecture} to map the concept of \textit{tie strength}~\cite{viswanath09evolution,wilson09user}. This concept is key to understanding how networks grow over time. One of the most influential publications on the topic, by the sociologist Mark Granovetter~\cite{Granovetter1973strength}, showed that strong ties tend to form close-knit clusters, while weak ties bridge across clusters. This fundamental work laid the foundation for an entire field of research. Sociologist Ronald Burt~\cite{burt2009structural} built upon that work by introducing the concept of structural holes to better explain how ideas spread: new ideas are likely to come from weak ties created by gatekeepers who bridge different communities. The notion of tie strength is also connected to how personal communities form. In studying how people grow their social circles~\cite{dunbar1995social,hill2003social}, anthropologist Robin Dunbar found that people perceive their social circles as relationships arranged by importance (i.e., by tie strength)~\cite{zhou2005discrete}. A typical social circle can be represented as an onion with five layers (relationship types) of decreasing intimacy: the support clique (at the core), the sympathy circle, close relationships, stable relationships (usually around 150 people, also known as the Dunbar Number~\cite{dunbar1998social}), and acquaintances. Granovetter's work has been further used to: i) predict tie strength from Facebook, Twitter, and question-answering data in the field of HCI~\cite{gilbert09predicting,gilbert2012predicting,panovich2012tie}; ii) study community structures in social media~\cite{grabowicz2012social,borregon14characterization,aiello15group}; and iii) study signed edges---positive \emph{vs.} negative edges~\cite{kunegis09slashdot,kunegis13what,Leskovecetal:2010}---to test the ``structural balance'' theory in digital networks, which suggests that ``the enemy of my enemy is my friend''.

This research makes clear that tie strength is an important dimension along which social relationships can be categorized, but also calls attention to the possibility that there might be other orthogonal dimensions that are also important to understanding relationships. For example, it is easy to think of relationships which are similar in tie strength (\emph{parents vs. romantic partner}) yet vary drastically in nature.

\subsection{Online conversations}

In interaction networks, a social link can be annotated with the messages that the two endpoints exchange. The analysis of textual exchanges yields rich information on the people involved in those conversations, and on what their relationship is about.

Online conversations on social media have been investigated extensively by computer scientists, largely because being able to automatically detect conversations that are interesting~\cite{dechoudhury09conversations} or relevant to key topics~\cite{becker2011beyond,aiello2013sensing} is a powerful tool to foster user engagement~\cite{harper07talk}. By mining conversations from emails and Twitter data, for example, researchers have identified the emergence of conventions (e.g., the adoption of specific language markers)~\cite{honey09beyond,boyd10tweet,kooti15evolution} and regularities in terms of thread length and number of participants~\cite{kumar10dynamics,backstrom2013}.

Researchers have been able to accurately predict personality traits from online conversations~\cite{mairesse2007using,quercia2011our}. Correa \emph{et al.}~\cite{correa10who} found that openness to new experiences and emotional stability correlate with the propensity to start new conversations online. Similarly, from tweets, Celli and Rossi~\cite{celli2012role} estimated user's emotional stability and  propensity to start new conversations.

Other research has investigated the topics discussed in conversations~\cite{purohit14understanding,inches2011online}. For example, Bearman and Parigi~\cite{bearman2004cloning} analyzed data from the GSS to learn what are the ``important matters'' people usually talk about. They manually identified 9 recurring macro-topics: community issues, news, kids, politics, health, relationships, money, ideology, and work. Meanwhile, Java \emph{et al.}~\cite{java07why} manually classified tweets into informational, conversational, and tweets for casual chatter. The availability of accurate sentiment analyzers able to work on short text snippets~\cite{goncalves13comparing} allowed for the exploration of emotions in dialogs. Kim \emph{et al.}~\cite{kim12feel} extracted topics from Twitter conversations with LDA, used Plutchik's model~\cite{plutchik80emotion} to assign emotions to them, and looked at emotional arcs  in conversations through time.

As opposed to this previous research, our work aims at identifying aspects that are more related to the \textit{communicative acts} of a conversation rather than its \textit{semantics}. While semantics focuses on the topics people talk about, communicative acts shape the actual nature of a relationship as well as both party's perceptions of that relationship. For example, we use language to perform a variety of social functions such as giving comfort and apologizing.

\subsection{Multiplex networks}

Research in temporal, dynamic, and multilayer networks explores different ways to represent social relationships.  In those areas, excellent introductions include manuscripts by Holme~\emph{et al.}~\cite{holme12temporal}, Gautreau~\emph{et al.}~\cite{gautreau09microdynamics}, and Kivela~\emph{et al.}~\cite{kivela14multilayer}. Scientists in the field of complex systems have modeled multi-relational social structures to account for the different kinds of social ties each individual has. Theoretical models were used to show how multiplex structures can influence collective dynamics such as cooperation and emergence of consensus~\cite{gomez2012evolution}. Few went beyond theoretical models and studied multiplex networks in real-world contexts. Szell~\emph{et al.}~\cite{szell2010multirelational} studied an interaction network from a massive multiplayer online game where relationships were characterized along the dimensions of friendship, communication, trade, enmity, aggression, and punishment.

Research in this area has focused on modeling a social graph's statistical properties but, in so doing, has  assumed the existence of certain kinds of relationships without specifying a comprehensive list of them. Here we focus on determining the most important types of real-world relationships.

\subsection{Multidimensional models in Sociology}\label{sec:related:multidimensional}

Many substantive social network studies represent ties with attributes other than friendship or acquaintanceship, including mentorship~\cite{podolny1997resources}, advice~\cite{gibbons2004friendship}, romance~\cite{parks1983romantic}, identity~\cite{white2008identity,tajfel2010social}, and emotional support~\cite{van1993delineating,walker1977social}. A few researchers have also attempted to draw a comprehensive characterization of the multiple sociological dimensions that describe a social relationship. Using the expression ``colors of closeness'', sociologist Barry Wellman has argued that the concept of tie strength is not sufficient to grasp a relationship's subtleties~\cite{rainie2012networked}. Similarly,
Bicchieri~\cite{bicchieri06grammar} and DeDeo~\cite{dedeo2012collective} expressed the need for identifying fundamental blocks of social interactions---the ``grammar of society'', in Bicchieri's words. 

In the 90s, Alan Page Fiske first proposed four basic types of sociality ~\cite{fiske1992four}, which formed the basis of his relational models theory. At around the same time, Wellman and Wortley~\cite{wellman90different} published the first work that studied how different types of ties result in different kinds of social support. By interviewing 29 people in Toronto, they found that five main dimensions of support---emotional aid, small services, financial aid, large services, companionship---are indeed provided by very different types of relationships (e.g., parents, spouse, friends). More than a decade after, Spencer and Pahl~\cite{spencer2006rethinking} conducted in-depth interviews to learn how people's ``personal communities'' are structured. Besides identifying the strength of ties (high/low commitment relationships), interviewees portrayed a number of stereotypical relationship types: associates, useful contacts, fun friends, favor friends, comforters, confidants, help-mates and soul-mates.

The multidimensional nature of social ties has been studied in the online context too. Using Emerson's and Blau's Social Exchange Theory~\cite{emerson1976social,blau64exchange}, Aiello \emph{et al.}~\cite{aiello14reading} found that individuals tend to exchange three types of what that theory would call ``resources'': knowledge, social support, and status. Status and power---two complementary aspects of the same type of social exchange---have been associated with the use of linguistic styles~\cite{bramsen11extracting,danescu12echoes,tchokni14emoticons}, and with different outcomes in online discussions~\cite{purohit14understanding}. Recently, Purohit~\emph{et al.} investigated the relationship between social cohesion, social identity and discussion divergence in online groups~\cite{purohit14understanding}, and discussed the power dynamics which result from them~\cite{tchokni14emoticons}. Budak~\emph{et al.} identified the presence of emotional support and information exchange in Twitter conversations, and discussed how those two domains were linked to a shared sense of community.

By proposing a categorization of fundamental types of social interactions, our work builds on this past research, which we elaborate on in \S\ref{sec:discussion:theoretical}.

\subsection{Research questions}
In our attempt to create a taxonomy of social relationships, we set out to answer three main research questions:
\begin{description}
\item \textbf{RQ1:} What are the main dimensions along which research fields (e.g., social psychology, sociology) have studied social relationships? (\S\ref{sec:meth_lit})
\item \textbf{RQ2:} Can we identify a more comprehensive set of dimensions that correspond to how people actually perceive and categorize their own social relationships? (\S\ref{sec:meth_st1})
\item \textbf{RQ3:} Do the newly-found dimensions capture previously unseen aspects of a network's structure? (\S\ref{sec:linkprediction})
\end{description}

\section{Methods and Results}\label{sec:methods}

We now elaborate on the methods we used to answer our three primary research questions and our findings for each of those questions.

\subsection{Literature review (RQ1)}\label{sec:meth_lit}

\begin{figure}[!b]
\centering
\includegraphics[width=0.99\textwidth]{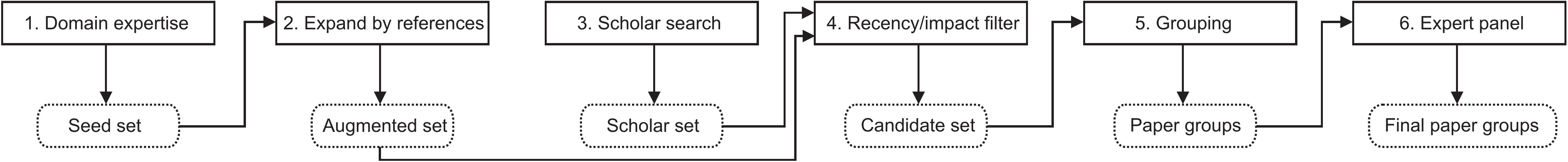}
\caption{6-step process for gathering the set of research papers about the core dimensions of social relationships.}
\label{fig:procedure}
\end{figure}

To find a set of relevant academic articles on the categorization of social relationships, we used an iterative procedure that mixed searches through online corpora with selections from domain experts. The process, similar to techniques used in the past to perform literature reviews in other domains~\cite{dittus2017community}, is represented in Figure~\ref{fig:procedure}.

To begin, we selected a few prominent seed papers chapters and books, which are well-known in several domains of research in the fields of sociology and social psychology~\cite{baumeister1995need}. When then expanded this set, with the help of Google Scholar, by adding papers referenced in and referencing by the seed papers. Independent from the initial seed set, we queried Google Scholar---which has been shown in prior studies to have comparable recall and precision to similar corpora~\cite{boeker2013google,bramer2013comparative,gehanno2013coverage,wright2014citation}---to get additional sources. We refined the search strategy iteratively to increase the recall of known sources and also refined the search keywords by reviewing the results returned. We structured the final query as follows:
{\small
\begin{verbatim}
intitle:(social OR group OR community OR network OR crowd
OR relationship OR tie OR link OR interaction OR conversation
OR friendship OR kinship OR interpersonal OR dyadic) AND
(type OR nature OR taxonomy OR classification OR categorization
OR model OR survey OR evolution OR exchange OR language)
\end{verbatim}
}

These papers were then categorized such that articles on the same topic were grouped together: for each article, we enumerated all the relevant keywords, clustered them by theme and designated a single word term to summarize each of these categories (e.g., similarity, homophily, and assortative matching were grouped together in the ``Similarity'' cluster). This grouping was done through unanimous consensus of all the authors. To avoid this leading to notable exclusions, five external experts in social psychology and sociology looked through these articles and augmented the list if there were any notable exclusions. Next, again through unanimous consensus, redundant articles (e.g., incremental wrt highly influential articles) were removed such that only the original influential article or articles in that research line were retained.

{\def\arraystretch{2}
\begin{table*}[ht]
\caption{Dimensions characteristic of social relationship as they emerge from the literature review.}
\centering
\begin{tabular}{p{38mm} p{30mm} p{37mm} p{20mm}}
\Xhline{2\arrayrulewidth}
\textbf{Characteristics} & \textbf{Simple Term} & \textbf{Disciplines} & \textbf{Sources} \\  
\hline

Similarity, Homophily,  \break Assortive Matching & Similarity & Sociology Psychology & \cite{mcpherson2001birds}, \cite{jackson2010social} \\
Need to Belong, Warmth, Social Support & Social Support & Psychology &  \cite{baumeister1995need}, \cite{fiske2007universal}, \cite{vaux1988social}\\
Trust, Reliability &Trust & Psychology Sociology Economics Biology &  \cite{luhmann1982trust}, \cite{zaheer1998does}\\
Power, Resources & Power & Sociology Psychology & \cite{french1959bases}, \cite{french1956formal}, \cite{blau64exchange} \\
Knowledge, Competence & Knowledge Transfer & Psychology  & \cite{fiske2007universal}, \cite{levin2004strength} \\
Group Membership, \break Culture, Identity & Identity & Sociology Psychology & \cite{tajfel2010social}, \cite{oakes1994stereotyping}, \cite{cantor1979prototypes} \\
Status, Respect & Status & Sociology Psychology & \cite{blau64exchange}, \cite{emerson1976social} \\
Mating, Sex, Romance & Romance & Psychology Sociology Economics Biology & \cite{buss2003evolution}, \cite{buss1993sexual}, \cite{emlen1977ecology} \\
\Xhline{2\arrayrulewidth}
\end{tabular}
\label{table:literature_review}
\end{table*}
}


This left us with a list of 19 influential articles (Table~\ref{table:literature_review}) that define eight broad dimensions, which we now describe:

\vspace{4pt}\noindent\textbf{Similarity.} This describes a relationship in which the two parties occupy a similar station in life. Phrased another way, this dimension describes the spatial closeness of two people in a highly dimensional demographic space. For example, two people would be high on this dimension if they are are of  similar ages, similar genders, work in similar industries, and earn similar salaries. Meanwhile, two people would be low  if they are from different age cohorts, are of opposite genders, and occupy different economic brackets~\cite{mcpherson2001birds,jackson2010social}.

\vspace{4pt}\noindent\textbf{Identity.} A relationship in which two people are brought together by their shared sense of belonging to a community that is personally meaningful to them and forms of a basis of their sense of self. One example might be two people united by virtue of being lifelong fans of the same sports team, another might be members of a marginalized racial or ethnic group (while these people are likely to be high in similarity as well, this is an orthogonal relationship dimension as it corresponds to a psychological identification with that group)~\cite{hastorf1954they, tajfel2010social,oakes1994stereotyping,cantor1979prototypes}.

\vspace{4pt}\noindent\textbf{Knowledge.} A relationship in which the exchange of information is a focal point of the relationship. This exchange may be asymmetric or not. That is, the information transfer may be largely unidirectional (e.g., the relationship between a math professor and his students, in which, it is mostly the professor giving the students information). Or the information transfer may be bidirectional (e.g., two CEOs in different industries who regularly call each other to  exchange information about the state of affairs in each others' industries)~\cite{fiske2007universal,levin2004strength}.

\vspace{4pt}\noindent\textbf{Social support.} A relationship in which one or both parties provide some form of aid to the other. This aid might come in several different forms. Some major categories of aid include: emotional aid, small favors (e.g., lending household items), long term services (e.g., regular help with health issue), financial aid, and companionship~\cite{wellman90different,baumeister1995need,fiske2007universal,vaux1988social}.

\vspace{4pt}\noindent\textbf{Trust.} This characterizes a relationship between two parties where one party is willing to rely on the other. This usually involves one person willfully allowing their fortune to be dictated by the actions of the other (this dynamic is captured in the economic game of the same name, the ``trust game'', in which one person may choose to allocate money to another, whereupon the sum is increased, but the final allocation of the sum is determined by this second person to whom the sum has been handed)~\cite{luhmann1982trust,zaheer1998does}.

\vspace{4pt}\noindent\textbf{Power.} A relationship in which one party has more resources than another or the ability to influence or control that party's behavior (to some extent) regardless of that party's willingness. Some classic examples would include the relationship between an autocratic ruler and his or her subjects or between a boss and his or her employees---both of which are characterized by a disparity in resources and the ability of one party to influence and control the behavior and outcomes of the other~\cite{french1959bases,french1956formal,blau64exchange}.

\vspace{4pt}\noindent\textbf{Respect.} A relationship in which one party confers status upon the other. An example may be the relationship between a judge and a law clerk, from the perspective of the law clerk~\cite{blau64exchange,emerson1976social}.

\vspace{4pt}\noindent\textbf{Romance.} A relationship characterized by intimacy goals. That is, the parties who are sexually interested in each other and or see each other as long term partners. Some relationships described by this dimension would include: a couple who are dating or two parents who share a life and are raising a kid together ~\cite{buss2003evolution,buss1993sexual,emlen1977ecology}.

\subsection{Crowdsourcing studies (RQ2)}\label{sec:meth_st1}

Following this literature review which identified eight key preliminary relationship dimensions, we conducted two empirical studies on Mechanical Turk (MTurk). The goal of the first study was to have people describe their relationships in everyday language and then to use these descriptions to generate a set of concepts that people associate with their relationships. The goal of the second study was to narrow down and cluster these concepts according to similarity and then to check the extent to which these clusters match the dimensions identified in the literature review as well as examine whether additional concepts should be added to the literature-driven list. Across the two studies, we recruited a total of 200 workers ($45\%$ women, $81\%$ White from Canada, United Kingdom and United States).

In the first study, to gain an understanding of how people perceive and describe their relationships in natural, ordinary language, we asked $N=100$ MTurk users to pick up to 3 single words that ``best describe'' their relationships and up to 3 words that indicate what ``matter most'' for their relationships. They were then asked to pick two words that best described their relationship with each of these four type of bonds: friends, romantic partners, coworkers, and parents. We asked about generic relationship first to avoid to biasing participants' responses by immediately focusing them on specific relationships types. However, to also allow participants to provide descriptors for specific relationships in their lives, we repeated the question for the four relationship roles listed above. Thus, we were able to collect and incorporate in our analyses both information about descriptors of people's relations in general and descriptors of their specific relationships.

\begin{figure}[t!]
\centering
\includegraphics[width=.70\textwidth]{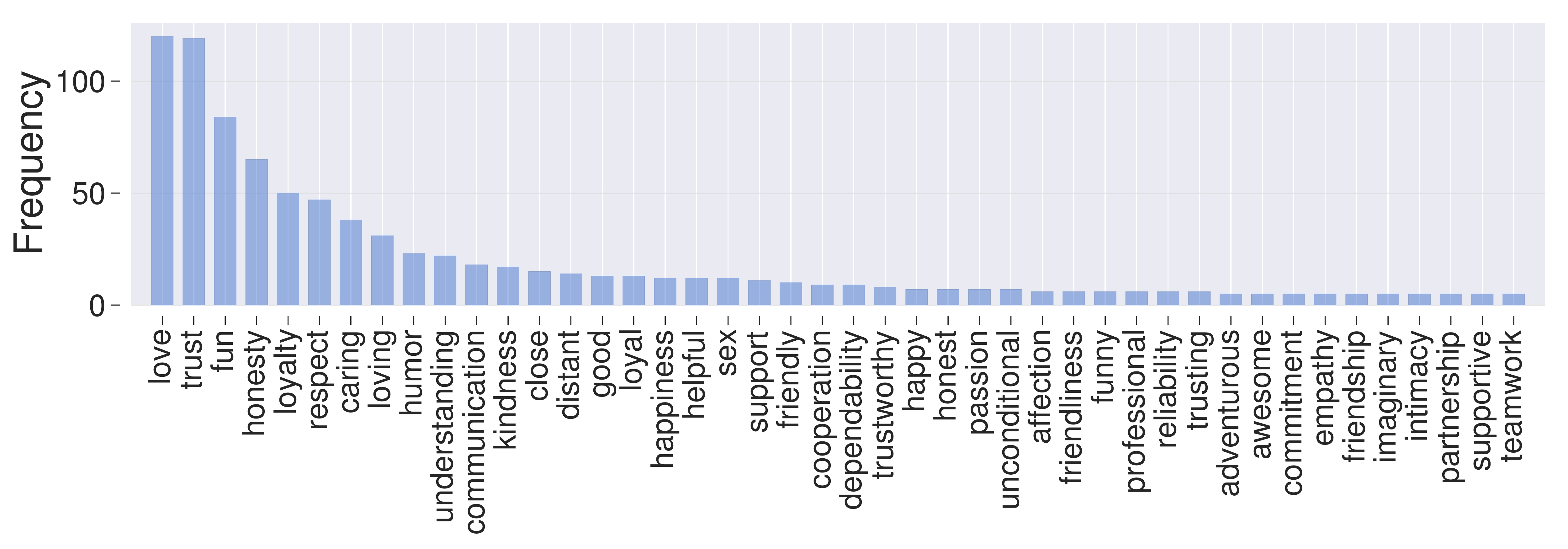}
\caption{Frequency of the top terms among the 220 unique terms used by the MTurk users to describe their social relationships.}
\vspace{-5pt}
\label{fig:term_frequency}
\end{figure}

In total, each user was supposed to provide 14 words (although some missed a few). Across all users we collected $1,352$ words (e.g., love, trust, empathy). After redundant words were eliminated, we were left with a final set of $220$ unique words that people use to describe their social relationships (the most frequent terms are shown in Figure~\ref{fig:term_frequency}).

This data was used as the basis of a second study aimed at validating and expanding the dimensions identified in the literature. A new set of $N=100$ MTurk users were each asked to rate every one of the 220 unique words in terms of how well those words described the relationships in general (from 1 = not at all, to 5 = very well). To ease the rating task, the words were shown to each participant in a grid format where each row in the grid corresponds to one of the 220 words, and there is a column for each of the scale values. The median time it took for users to rate every word was 8.3 minutes. We compensated people at a rate that was at least as high as the U.S. minimum hourly wage. The order in which each of the words appeared was randomized between participants.

As a result, each word $w$ can be 
characterized by a 100-dimensional rating vector $\overline{w}$ whose $u^{th}$ entry $w_{u}$ contains the rating given to it by user $u$. These vectors can then detect ``implicit mental associations'' and, as such, they go beyond semantic similarity. As a concrete example, two words with similar vectors (e.g., ``openness'' and ``patience'') are not necessarily semantically similar but people tend to mentally associate them (e.g., as revealed by a high correlation between the ``openness'' and ``patience'' vectors). Consequently, by clustering words based on their vectors, we can obtain clusters of words that reflect how people tend to think about relationships. 

\begin{figure}[t!]
\centering
\includegraphics[width=.80\textwidth]{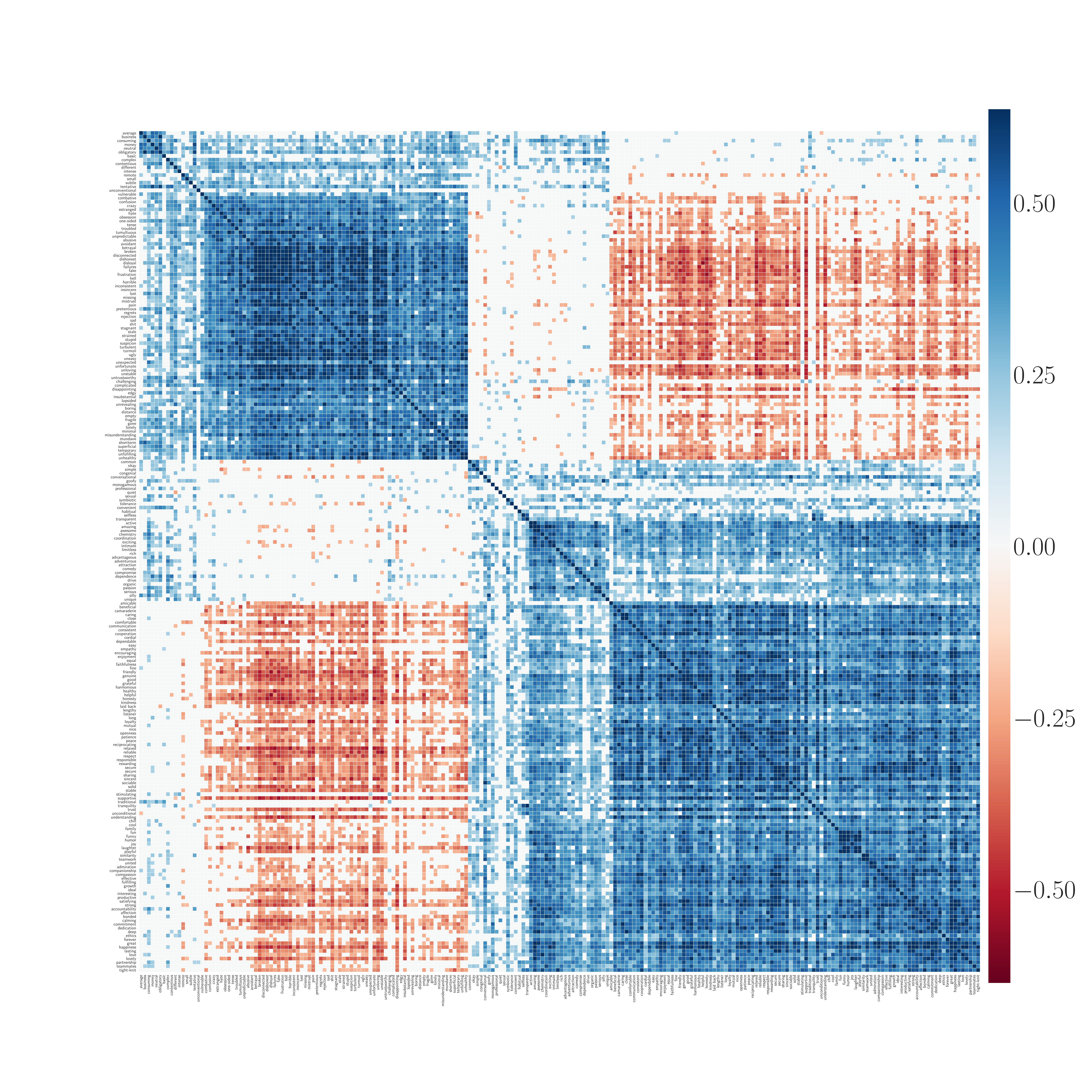}
\caption{Correlation matrix of words that describe social relationships. Rows are sorted according to the blockmodeling output. This image has been created at high resolution to allow for the inspection of individual words by zooming on screen.}
\vspace{-5pt}
\label{fig:blockmodeling}
\end{figure}

More specifically, we first computed the Spearman rank correlation coefficient between the vectors $\overline{w}$ of all the word pairs (analogous results are obtained when using the Pearson coefficient). This step yielded a symmetric $220 \times 220$ correlation matrix $M$ whose generic entry $M_{ij}$ contains the correlation score between words $i$ and $j$. To extract groups of related words, we run blockmodeling on the matrix $M$ to group together words that exhibit similar patterns. The matrix is partitioned hierarchically in exactly two clusters at each step. The procedure is repeated recursively on the sub-matrices until clusters contain one single word. To select the optimal level in the hierarchy, we measured the average rank correlation between the vectors $\overline{w}$ of all the members of every cluster: if the average correlation is lower than the one found in the previous level, the recursion stops. Figure~\ref{fig:blockmodeling} shows the final matrix $M_{ij}$ with the rows and columns re-arranged so that words in the same cluster are listed in adjacent rows.

\begin{figure}[t!]
\centering
\includegraphics[width=.50\textwidth]{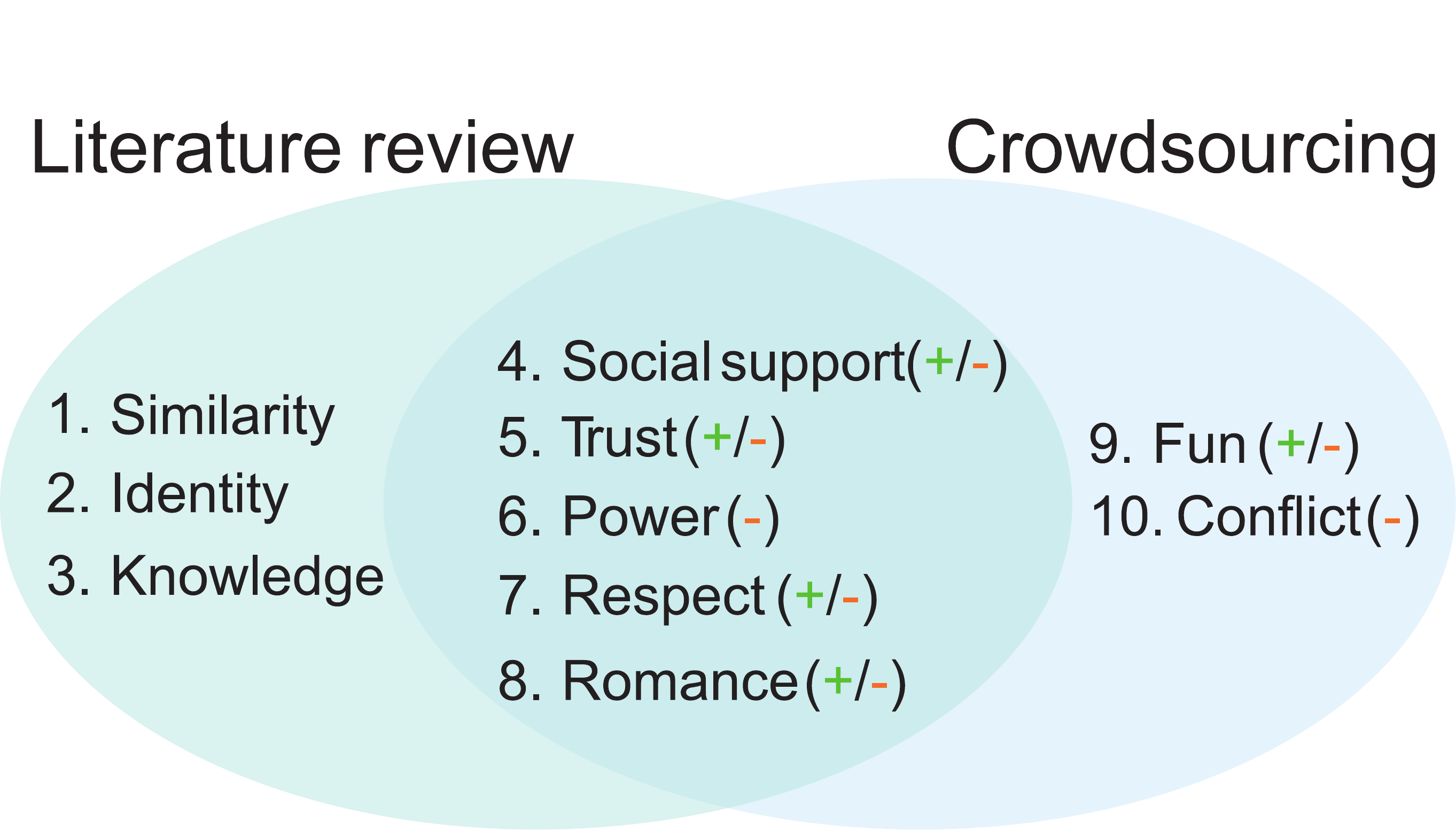}
\caption{Ten dimensions from literature and experiments. Five appear in both literature and experiments (middle), three appear in literature only (left), and two in the study only (right). Negative and positive signs correspond to valence; dimensions with both signs may be positive or negative in valence depending on the presence or absence of that dimension.}
\vspace{-10pt}
\label{fig:venn}
\end{figure}

The first binary partition yields a clear-cut dichotomy between positive and negative words. The positive words are best grouped into five leaf-categories. We found one category corresponding to romantic relationships, one that encompasses the notion of social/emotional support, one expressing trust, and one containing indicators of status and respect (e.g., ``admiration''). The fifth category  did not emerge in our literature review and consists of descriptors such as laughter, joy, and humor (``fun,'' for brevity). The negative words are best grouped into seven categories.  Five of them reflect the lack of social support, romance (``unloving''), trust (``untrustworthy''), respect (``disappointing'', ``insubstantial''), and fun (``boring''). The sixth category refers to power, which our literature review already uncovered (e.g., relationships that are ``obligatory'' or exist because of some monetary dependency). The seventh and final category did not emerge in our literature review and can be described as something along the lines of conflict and hatred. The resulting 10 dimensions and related terms are made publicly available for further research use\footnote{\url{www.tinghy.org/data}}. Figure~\ref{fig:venn} provides a summary. Three categories (i.e., identity, similarity, and knowledge transfer) are discussed in the literature but they do not emerge from the user study, not least because they are the hardest to verbalize. The other two categories of  `having fun together' and conflict, instead, are not explicitly delineated in the literature\footnote{As we shall discuss in \S\ref{sec:discussion:theoretical}, fun is mentioned by Spencer and Pahl in their relationship categorization~\cite{spencer2006rethinking} but it is not analysed as a sociological concept in any of the papers that we have reviewed.} but tended to be more frequently verbalized. The remaining five dimensions are reported both in the crowdsourcing study and in the literature.

To check whether the results of the crowdsourcing experiments varied along demographic lines, we conducted separate splits of the responses by gender (male, female), age (above or below median), and racial background (white, and non-white as these were the two largest bins). For each demographic split, there is always a high correlation between responses of the two groups. In the first study, the correlation between the frequencies of the words fluctuates in the range [0.83-0.92]. Likewise, in the second study, the correlation between the values in the word-word correlation matrix computed across splits always ranges between [0.72-0.79].

\subsection{Relationship dimensions for link prediction (RQ3)} \label{sec:linkprediction}

To show a concrete application of these relationship categories that we derived, we ran a link prediction experiment. Link prediction is a fundamental task in network science that aims to anticipate which edges will appear in an evolving graph~\cite{aiello12friendship}. In social networks, the ability to reliably predict new social connections has important theoretical implications because it improves our ability to explain the dynamics that drive the evolution of ego-networks and, more generally, of social systems. In social media, it also serves the practical purpose of providing recommendations to users about new potential social connections they could establish. In large systems, even a slight improvement of the accuracy of these recommendations triggers the creation of a very large number of additional social connections. In the following, we measure how incorporating the notion of the newly found social dimensions into a link prediction model can lead to such an improvement compared to traditional approaches based on structural indicators only.

We used a complete snapshot from aNobii.com---a social platform for book lovers---which was collected in 2012 and made available for research purposes~\cite{aiello2012link}. On aNobii, users create their own digital libraries containing books they have read and wish to read in the future. They connect with others through directional social links that might or might not be reciprocated, similar to the Twitter's ``follow'' relation. Any pair of users can communicate by writing messages on each other's public profile. Rather than looking at the follower network, we focused on the communication network: a graph where nodes are users, and a directed link between nodes is drawn if the origin node has sent at least one message to the destination node. We annotated every link in the communication network with the bag of words of all the messages that have been sent on that link. Overall, the full communication network contains about 80k nodes and 575k links.

The classical formulation of the link prediction problem in a static scenario is to consider a snapshot of the network and predict whether a link between two nodes exists or not based on a number of features that describe the two nodes or the context around them~\cite{liben2007link}. The task can be solved using a wide range of techniques~\cite{lu2011link} but, in practice, the majority of the local features that end up being predictive tend to often be simple variations of the common neighbors metric: that is, the probability of a social connection between two individuals grows with the number of contacts they have in common. Also, those metrics are estimators of the strength of the (potential) tie between two people. In this experiment we use \textit{triangle overlap} ($TO$) as a baseline feature:
\begin{equation}
TO(u,v) = \frac{\Gamma_{out}(u) \cap \Gamma_{in}(v)}{|\Gamma_{out}(u)|}
\end{equation}
where $u$ and $v$ are two users, and $\Gamma_{in}$ and $\Gamma_{out}$ are, respectively, their sets of in- and out-neighbors. In short, $TO$ is a version of common neighbors adapted to directed networks and normalized by the size of $u$'s neighbor set.  In aNobii, $TO$ has been found to perform better than all the other structural metrics (including simple common neighbors) and than semantic features~\cite{aiello2012link} (e.g., similarity of user profiles based on their digital bookshelves).

In the context of this work, to exploit our newly formulated ten dimensions, instead of considering the common neighbors as homogeneous, we can partition them according to the types of relationships  they are involved in. To do that, we matched the crowd-sourced terms that reflect the concepts of support, trust, power, respect, romance, fun, and conflict (Figure~\ref{fig:venn}) with the bag of words of each social link. We do not use the dimensions of similarity, identity, and knowledge because the crowdsourcing did not generate any words for those three. We then label the link with the dimension having the highest number of matching words. As a result, for a user pair $(u, v)$, we obtain a social dimension vector $D(u,v)$ whose entries count the number of common neighbors who are connected to $u$ with a link of a given type (e.g., 2 common neighbors of type ``support'', 1 of type ``respect''). $D(u,v)$ is used as feature set for our approach.

\begin{table*}[b!]
	\centering
	\caption{Link prediction results on 200k links using as features i) triangle overlap, ii) the vector of relationship dimensions and iii) those two combined. Percentage increase over triangle overlap is shown.}
	\begin{tabular}{c|cc}
		\textbf{Features}    & \textbf{Precision} & \textbf{AUC} \\
		\hline
		Triangle overlap  & 0.749 & 0.755 \\
		Relationship dimensions   & 0.800 (+7\%) & 0.803 (+6\%) \\
		All        & 0.783 (+5\%) & 0.841 (+11\%) \\
	\end{tabular}
	\label{tab:linkprediction}
\end{table*}

At random, we selected 100k connected user pairs (positive examples) and 100k disconnected ones at 2 hops from each other (negative ones), and attach to each pair the two predictors $TO$ and $D(u,v)$. Upon 10-fold cross validation, a Random Forest classifier with $D(u,v)$ as predictor brings an improvement of 7\% in AUC, and an improvement of 6\% in accuracy compared to a Random Forest classifier based on $TO$ (Table~\ref{tab:linkprediction}). When combined, the two predictors yield an increase of 11\% in AUC. In large networks such as ours, even an improvement of 1\%, apparently small in itself, is crucial to capture a considerable part of the network structure that would be otherwise invisible to the method of triangle overlap, which has been previously found to be the best performing method~\cite{aiello2012link}. 

To assess the predictive power of the different social dimensions in $D(u,v)$, we measured the relative feature importance in the Random Forest classifier (a score that sums to 1 over all the dimensions). We found that social support (0.32) and respect (0.29) are the most predictive features, followed by trust (0.18) and fun (0.12). Power (0.03), romance (0.03), and conflict (0.03) seem not to be very predictive, in this context. It is hard to extract generalizable findings from this feature ranking, though. In fact, labeling a relationship with the dimension whose associated words appear most frequently in the conversation is a crude approach and it may suffer from a number of limitations. Although it is not unreasonable that the words that people exchange bear a relationship to the type of tie people have, those words might not be always representative of the social dimensions they are associated with, not least because their semantics changes depending on the context of the conversation. Even if they were representative, these results might not be generalizable across platforms: the link creation dynamics of an interest-based network tailored for book lovers are arguably different from other types of social media. In this experiment, we use word matching to provide a simple proof-of-concept that decomposing an indicator that reflects the number of common neighbors into a vector of relationship types increases the link prediction power. In the discussion that follows, we sketch additional methods that could be used to collect data for training supervised models and even more accurately map conversations to social dimensions.

\section{Discussion} \label{sec:discussion}

Our work has aimed at  understanding how individuals conceptualize social relationships, and the results provide a basis to move forward with social network analyses in more sophisticated ways. By categorizing relationships according to sociologically and psychologically meaningful variables, we are likely to be able to understand and predict more of both people's offline and online behavior. For example, in modeling people's behavior, for two given links with the same weight, it might be very useful to know the ``type'' of link for each. For a given person, the link weight between them and their co-worker might be similar to the weight with their friend, if weights represent something like frequency of interactions. However, knowing that one of these links is primarily characterized by power and status while the other is primarily characterized by trust and fun, is likely to be highly informative and predictive above and beyond the basic information.

\subsection{Theoretical implications} \label{sec:discussion:theoretical}

{\def\arraystretch{2}
\begin{table*}[t!]
\small
\centering
\caption{Comparison between our set of dimensions and the concepts presented in the two multidimensional models of social interactions by Spencer \& Pahl and Wellman \& Wortley}
\begin{tabular}{p{22mm} p{46mm} p{46mm}}
\Xhline{2\arrayrulewidth}
\textbf{Our model} & \textbf{Spencer \& Pahl}~\cite{spencer2006rethinking} & \textbf{Wellman \& Wortley}~\cite{wellman90different}\\
\hline
\textit{Similarity}     & \textit{Associates} (share a common activity or hobby) & - \\
\textit{Social support} & \textit{Comforters} (providing deep level of emotional support) & \textit{Emotional aid} (advice, minor or major emotional aid) \\
\textit{Trust}          & \textit{Confidants} (disclosure of personal information) & \textit{Large services} (major services including health care, child care)\\
\textit{Power}          & - & - \\
\textit{Knowledge}      & \textit{Useful contacts} (share information and advice, typically related to work) & \textit{Companionship} (discussing ideas) \\
\textit{Identity}       & - & \textit{Companionship} (participating in common groups) \\
\textit{Respect}        & - & - \\
\textit{Romance }       & - & - \\
\textit{Fun}            & \textit{Fun friends} (socialization primarily to have fun) & - \\
\textit{Conflict}       & - & - \\
-              & \textit{Favor} (providing functional help) & \textit{Small services, financial aid} (lending items or money, small aid) \\
\Xhline{2\arrayrulewidth}
\end{tabular}
\label{tab:comparison}
\end{table*}
}

Our work makes two main contributions to theory. First, while not dispositive, we highlight and provide an empirical basis for several important dimensions along which social relationships may be conceptualized. Second, these dimensions offer a new way of modeling social networks, which goes beyond tie strength and relationship semantics. Of course, there have been efforts in the social sciences to provide meaningful categories for social relationships. For example, two prominent ones are outlined in Table~\ref{tab:comparison} (Spencer \& Pahl~\cite{spencer2006rethinking} and Wellman \& Wortley~\cite{wellman90different} (already mentioned in \S\ref{sec:related:multidimensional}). There is overlap between the dimensions outlined by these two theories and our categories, as there should be if all parties are studying the same underlying phenomenon. However, we believe that our categories emcompass a broader range of underling dimensions than either of these previous two theories---and further come with empirical validation as well as demonstrated usefulness in a concrete network analysis task (link prediction).

In the future, having tools to detect the type of tie will open up new questions which are meaningful theoretically and testable empirically. For example, which types of ties are prevalent in which social circles? What is the relationship between network structure and type of tie? Which ties are bonding and which are bridging?

\subsection{Practical implications} \label{sec:discussion:practical}

By categorizing relationships according to sociologically and psychologically meaningful variables, we are likely to be able to better understand and predict people's offline and online behavior. We have shown how using the information about the type of social relationship yields an improvement on prediction tasks on social media compared to using tie strength.

Moving forward, the relationship dimensions we identified could be useful in the design of social networking systems. We see at least three domains where online social media platforms could benefit from a system that is able to accurately estimate social relationship types:

\begin{itemize}

\item \emph{Managing privacy.} Privacy controls benefit from understanding the meaning of a tie. When users make privacy choices, a system could control which friends (e.g., social supportive friends \emph{vs.} knowledgeable ones) should be able to see which types of content (e.g., sensitive photos \emph{vs.} news articles).  That works for new users too: one could simply label the newly acquired relationship using our ten dimensions. As opposed to information sorting approaches based on tie strength,  those based on our labels are potentially able to appropriately place different types of information in  different social circles.

\item \emph{Curating news feeds.} People are turning to social media to satisfy social and informational needs: consuming news, forming opinions, and sharing problems with others are all activities that increasingly take place online~\cite{bakshy2015exposure,quan2010uses}. This trend calls for new solutions to curate news feeds that show the right information to the right friends~\cite{eslami2015feedvis,rader2015understanding}. Our fine-grained representation of relationships allows for a more targeted disclosure of such information, resulting in more useful and enjoyable news feeds. News from users who enjoy the trust of many people might be promoted in the feed over those from users who are popular but not necessarily trusted. Posts that express emotional distress could be shown first to friends who are able to provide emotional support.

\item \emph{Suggesting new friends.}  Social-networking sites often recommend new social contacts. Yet, we have good reasons to not befriend every single recommended user. A system that automatically (albeit partly) understands those reasons should know which are the building blocks of a stable social circle; its recommendations should not be blind to, for example, `power' and `conflict' dynamics which are accounted for in our taxonomy.

\end{itemize}

\subsection{Limitations and future work}\label{sec:discussion:tinghy}

\begin{figure}[!t]
\centering
\includegraphics[width=0.45\textwidth]{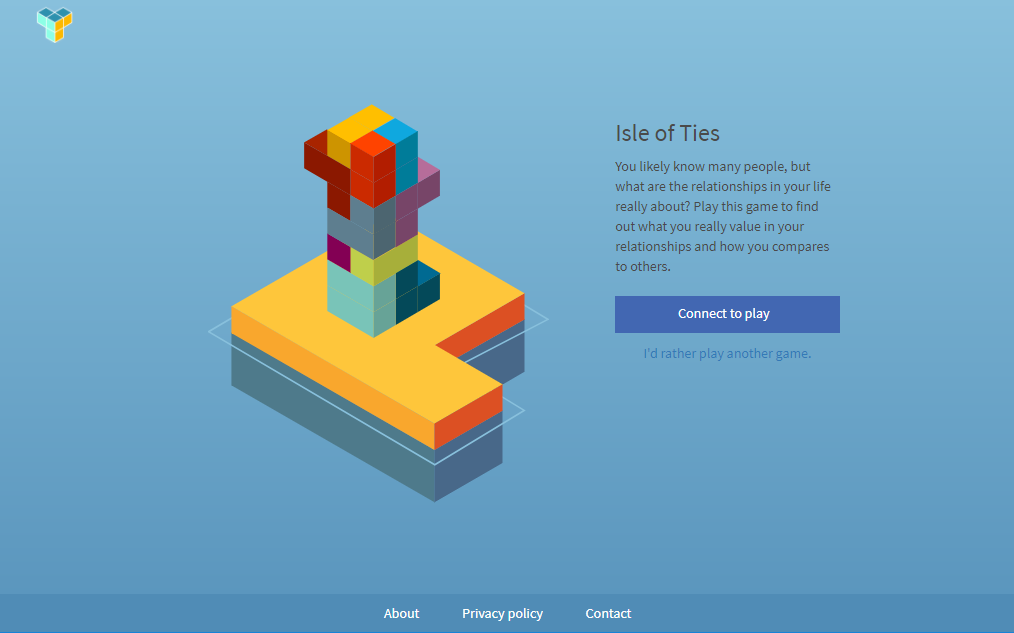}
\includegraphics[width=0.45\textwidth]{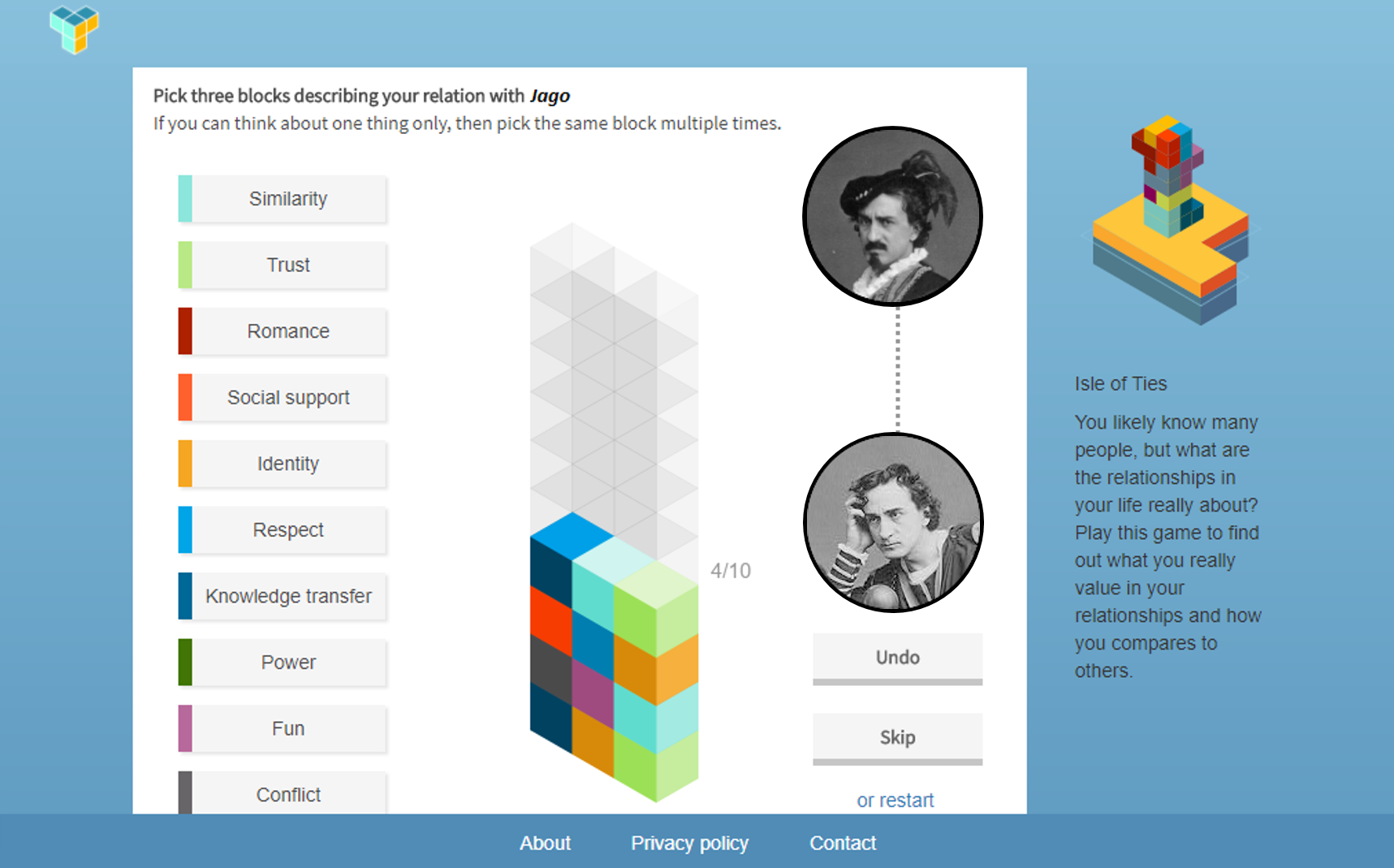}
\caption{Screenshots of the tinghy.org platform. Users log in with their social media accounts (left) and are asked to label the relationships with their friends using the ten dimensions we have identified in this study (right).}
\label{fig:tinghy}
\end{figure}

This work has two main limitations. The first concerns generalizability. Given that the majority of our crowd-sourcing respondents are White and from English-speaking countries, we should be cautious before generalizing our results across cultures---or even to certain subcultures within a country; thus, more cross-cultural work would be useful. Further research might also assess the extent to which prompting participants toward specific relationship roles (e.g., parents, co-workers) biased our results towards strong ties. The second limitation concerns the feasibility of collecting data. It might be difficult to train supervised algorithms to map relationships onto the types identified here, as it is unclear how to get users to label their relationships, and how to build  a reliable training set that goes beyond the simple word matching approach previously used in our link prediction task. To address this limitation, we developed \url{www.tinghy.org}, a web platform that makes the labeling task practical and fun. On this platform, participants can play a series of psychological games. In one of these games, users log in through their social media accounts, and they are sequentially presented with 10 of their actual friends. For each friend, they make choices which reveal the extent to which that relationship is described by our 10 dimensions (Figure~\ref{fig:tinghy}). The user interface is ``gamified'' so that the experience is fun and rewarding. As participants go through each of their friends, they build a wall representing the core components of their social circles. With the explicit consent of the user, interaction data between the players and their friends is gathered in background. This gamified approach is not platform-dependent and it can link data sources from different platforms that offer internet-mediated conversations. When users are logged in with Twitter, for example, the platform collects the text associated with mentions and replies exchanged with their contacts. This process results into a scalable way of collecting conversational text and its associations to categorized data about the type of the social relationships that mediate those conversations. We envision using the collected data to train supervised machine learning models that can predict the 10 dimensions from conversational text.

In the future, upon a large-scale collection of such labels, researchers could study the relationship between tie dimensions and a variety of factors derived from online data such as use of language, network structure, and real-world outcomes---building up a more robust understanding of the online markers of such relationship types.

\bibliographystyle{ACM-Reference-Format}
\bibliography{ms}

\end{document}